\documentstyle[preprint,aps,floats,epsfig]{revtex}
\def\erf{\,{\rm erf}\,}
\newcommand{\be}{\begin{equation}}
\newcommand{\ee}{\end{equation}}
\newcommand{\bea}{\begin{eqnarray}}
\newcommand{\eea}{\end{eqnarray}}
\newcommand{\bml}{\begin{mathletters}}
\newcommand{\eml}{\end{mathletters}}

\def\aprle{\buildrel < \over {_{\sim}}}
\def\aprge{\buildrel > \over {_{\sim}}}
\tighten
\begin{document}
\preprint{DTP/00/13, hep-th/0002072}
\draft

%\twocolumn[\hsize\textwidth\columnwidth\hsize\csname @twocolumnfalse\endcsname

%%%%%%%%%%%%%%%%%%%%%%%%%%%%%%%%%%%%%%%%%%%%%%%%%%%%%%%%%%%%%%%%%%%%%%%%%%

%\wideabs{                       % Uncomment this line for two-column output
\title{ 
Opening up extra dimensions at ultra-large scales
}
\author{Ruth Gregory$^{1}$, 
Valery A. Rubakov$^{2}$ and Sergei M. Sibiryakov$^{2}$  }

\address{~$^1$ Centre for Particle Theory, Durham University,
South Road, Durham, DH1 3LE, U.K.\\
{~}$^2$ Institute for Nuclear Research of the Russian Academy of Sciences,\\
60th October Anniversary prospect, 7a, Moscow 117312, Russia.
}
\maketitle

%%%%%%%%%%%%%%%%%%%%%%%%%%%%%%%%%%%%%%%%%%%%%%%%%%%%%%%%%%%%%%%%%%%%%%%%%%
\begin{abstract}
The standard picture of viable higher-dimensional
theories is that 
direct manifestations of
extra dimensions occur
at short
distances only, 
whereas long-distance physics is described by
effective four-dimensional theories.
We show that this is not necessarily true in
models with infinite extra dimensions. As an example, we consider a
five-dimensional scenario with three 3-branes in which gravity 
is five-dimensional both at short {\it and} very long distance scales, with
conventional four-dimensional gravity operating at intermediate length
scales. A phenomenologically acceptable range of 
validity of four-dimensional gravity extending from
microscopic to cosmological scales is obtained
without strong fine-tuning of parameters.
\end{abstract}
\pacs{PACS numbers: 04.50.+h, 11.25.Mj \hfill hep-th/0002072}
%}                               % Uncomment this line for two-column output

%%%%%%%%%%%%%%%%%%%%%%%%%%%%%%%%%%%%%%%%%%%%%%%%%%%%%%%%%%%%%%%%%%%%%%%%%%

Our world is often thought to have more than four
fundamental space-time dimensions.
This point of view is strongly supported by string/M-theory,
and higher dimensional
theories are currently being developed in various directions.
The standard lore is that in phenomenologically viable models,
extra dimensions open up at short distances only, 
whereas above a certain length scale, and  all the way up to infinite
distances, physics is described by effective four-dimensional theories.

In this letter we show that the latter picture is not universal:
there are models in which extra dimensions open up both at short
{\it and  very long} distances. Namely, gravity may become 
higher-dimensional in both of these extremes. At intermediate length
scales, physics is described by conventional four-dimensional laws, 
so the phenomenology of these models is still acceptable.

The starting point for our discussion is the observation  
\cite{Randall:1999vf} that extra dimensions may be infinite,
with the usual matter residing on a 3-brane embedded in 
higher-dimensional space. In the original Randall--Sundrum (RS) model
\cite{Randall:1999vf}, gravity is effectively four-dimensional
at large scales due to the existence of a graviton
bound state localized near the 3-brane in five dimensions.
We point out that in other five-dimensional models with an infinite
extra dimension (in particular, in the model introduced in
Ref.\cite{Charmousis:1999rg}), localization of the graviton may
be incomplete, and it is this  property that leads to the
restoration of the five-dimensional form of  gravity at very 
long distances. The latter 
feature is manifest both in the case of static
sources, where the four-dimensional Newton's gravity law changes to its
five-dimensional counterpart above a certain length scale, 
and in dynamics of gravitational waves
which disappear into the fifth dimension  thus violating
four-dimensional 
energy conservation, after travelling long enough
distance.
It is likely that these phenomena are not peculiar
to five-dimensional theories and occur
in a number of models with more than one extra dimension.

It has been found recently \cite{Kogan:1999}
that exotic large-distance effects  may also appear in models
with compact extra dimensions, due to the possible presence of
very light Kaluza--Klein states. This interesting
scenario is, however, considerably different from ours: in compact
models, extra dimensions show up at large distances somewhat
indirectly, through the spectrum of the corresponding four-dimensional
effective theory: four-dimensional energy is conserved, and so on.
In our case, physics at ultra-large distances in intrinsically
five-dimensional. 

As our concrete example, let us consider a five-dimensional
model of Ref.\ \cite{Charmousis:1999rg}.
The model
contains one brane
with tension $\sigma > 0$ and two branes with equal
tensions $-\sigma /2$
placed at equal distances 
to the right and to the left of the positive tension
brane in the fifth direction.
The two negative tension branes are introduced for simplicity,
to have  $Z_2$ symmetry, $z \to -z$, in analogy to
RS model (hereafter $z$ denotes the fifth coordinate). 
We assume that conventional matter resides on the
positive tension brane, and in what follows we will be
interested in gravitational interactions of this matter.
The $Z_2$ symmetry enables us to consider explicitly
only the region to the right of the positive tension brane.

The physical set-up (for $z>0$) is as follows:
The bulk cosmological constant between the branes, $\Lambda$, is
negative as in the RS model, however, in contrast to that model, is
{\it zero} to the right of the negative tension brane. With
appropriately tuned $\Lambda$,  there exists a
solution to the five-dimensional 
Einstein equations for which both positive and
negative tension branes are at rest at $z=0$ and $z=z_c$ respectively,
$z_c$ being an arbitrary constant. The metric of this solution is
\be
ds^2=a^2(z)\eta_{\mu\nu}dx^{\mu}dx^{\nu}-dz^2
\label{1}
\ee
where
\be
a(z) = \cases{ e^{-kz} & $0<z<z_c$ \cr
e^{-kz_c}\equiv a_- & $z>z_c$\cr}
\label{2}
\ee
The constant $k$ is related to $\sigma$ and $\Lambda$ as follows:
$\sigma=\frac{3k}{4\pi G_5}$, $\Lambda=-\sigma k$, where $G_5$ is
the five-dimensional Newton constant. The four-dimensional hypersurfaces
$z=const.$ are flat, the five-dimensional space-time is flat to the
right of the negative-tension brane and anti-de Sitter between the
branes. The spacetime to the left of the positive tension brane is
of course a mirror image of this set-up.

This background has two length scales, $k^{-1}$ and 
$\zeta_c \equiv k^{-1} e^{kz_c}$. We will consider the case of
large enough $z_c$, in which the two scales are 
well separated, $\zeta_c \gg k^{-1}$. We will see that
gravity in this model is effectively four-dimensional
at distances $r$ belonging to the interval
$k^{-1} \ll r \ll \zeta_c (k\zeta_c)^2$, and is five-dimensional
{\it both} at short distances, $r \ll k^{-1}$ (this situation is
exactly the same as in RS model), and at long distances,
$r\gg \zeta_c (k\zeta_c)^2$. In the latter r\'egime of very long
distances the five-dimensional gravitational constant gets effectively
renormalized and no longer coincides with $G_5$.

To find the gravity law experienced by matter residing on
the positive tension brane, let us study
gravitational perturbations about the background
metric (\ref{1}). We will work in the Gaussian Normal (GN) gauge,
$g_{zz}=-1$, $g_{z\mu}=0$.
The linearized theory is described by the metric
\be
ds^2=a^2(z)\eta_{\mu\nu}dx^{\mu}dx^{\nu}+
h_{\mu\nu}(x,z)dx^{\mu}dx^{\nu}-dz^2
\label{3}
\ee
There are two types of linearized excitations in this model.
One of them is a four-dimensional scalar --- the radion ---
that corresponds to the relative motion of the branes
(see, e.g., Refs. 
\cite{Csaki:1999mp,Goldberger:1999un,Charmousis:1999rg})
The wave function of the radion is localized between the branes,
and its interactions with matter on the positive tension brane
result in a scalar force of the Brans--Dicke type. If the distance
between the branes is not stabilized by one or another mechanism,
the radion has zero four-dimensional mass and gives rise to the usual
$1/r$ potential in an effective four-dimensional theory. In the 
particular model under discussion, the radion excitation has 
been studied in Ref. \cite{Charmousis:1999rg}. It is conceivable
that the distance between the branes may be stabilized (cf. Ref.
\cite{Goldberger:1999uk}); in which case the interactions due to
the radion switch off at large distances.

In this letter we are interested in other types of excitation
that leave the branes at rest, namely the five-dimensional gravitons.
When the radion is disregarded, there exists a frame which is 
GN with respect to both branes simultaneously.  In this frame, 
the transverse-traceless gauge can be chosen, $h^{\mu}_{\mu}=0$,
$h^{\nu}_{\mu,\nu}=0$, and the linearized Einstein equations take
one and the same simple form for all components of $h_{\mu\nu}$,
\be
\cases{ h'' - 4k^2 h - {1\over a^2} \Box^{(4)}h=0 & $0<z<z_c$\cr
h'' - {1\over a_-^2}\Box^{(4)}h=0 & $z>z_c$\cr}
\label{6*}
\ee
The Israel junction conditions on the branes are
\be
\cases{h'+2kh = 0 & at $z=0$ \cr \left[ h'\right] - 2kh = 0 & at $z=z_c$\cr}
\label{7*}
\ee
where $\left[h'\right]$ is the discontinuity of the
$z$-derivative of the metric perturbation at $z_c$, 
and four-dimensional indices
are omitted. A general perturbation is a
superposition of modes, $h=\psi(z)e^{ip_{\mu}x^{\mu}}$ 
with $p^2=m^2$, where $\psi$ obeys the following set of equations
in the bulk,
\be
\cases{ \psi'' -4 k^2\psi+\frac{m^2}{a^2}\psi = 0 & $0<z<z_c$ \cr
\psi''+\frac{m^2}{a_-^2}\psi=0 & $z > z_c$ \cr}
\label{9}
\ee
with the junction conditions (\ref{7*}) (replacing $h$ by $\psi$).
It is straightforward to check that there are no negative modes,
i.e., normalizable solutions to these equations with $m^2 < 0$.
There are no normalizable solutions with $m^2 \geq 0$ either, 
so the spectrum is continuous, beginning at $m^2 =0$. To write the
modes explicitly, it is convenient to introduce a new coordinate
between the branes, $\zeta=\frac{1}{k}e^{kz}$, in terms of which the
background metric is conformally flat. Then the modes
have the following form,
\be
\psi_m = \cases{ C_m\left[N_1\left(\frac{m}{k}\right)J_2(m\zeta)-
J_1\left(\frac{m}{k}\right)N_2(m\zeta)\right] & $0<z<z_c$\cr
A_m \cos\left(\frac{m}{a_-}(z-z_c)\right)+
B_m \sin\left(\frac{m}{a_-}(z-z_c)\right) & $z>z_c$ \cr}
\label{18}
\ee
where $N$ and $J$ are the Bessel functions.
The constants $A_m, B_m$ and $C_m$ obey two relations due to the
junction conditions at the negative tension brane. Explicitly, 
\bml\bea
A_m&=&C_m\left[N_1\left(\frac{m}{k}\right)J_2(m\zeta_c)
- J_1\left(\frac{m}{k}\right)N_2(m\zeta_c)\right]
\label{AA} \\
B_m&=&C_m \left[N_1\left(\frac{m}{k}\right)J_1(m\zeta_c) -
J_1\left(\frac{m}{k}\right)N_1(m\zeta_c) \right]
\label{BB}
\eea\eml
The remaining overall constant $C_m$ is obtained from
the normalization condition. The latter is determined by the
explicit form of Eq.\ (\ref{9}) and reads
\be
\int~\psi_m^{*}(z) \psi_{m'} (z) \frac{dz}{a^2(z)} = \delta (m-m')
\ee
One makes use of the asymptotic behaviour of $\psi_m$ at
$z \to \infty$ and finds
\be
\frac{\pi}{a_-}(|A_m|^2+|B_m|^2)=1
\ee
which fixes $C_m$  from (\ref{AA}) and (\ref{BB}).

It is instructive to consider two limiting cases. 
At $m\zeta_c\gg 1$ we obtain by making use of the
asymptotics of the Bessel functions,
\be
C_m^2=\frac{m}{2k}\left[J_1^2\left(\frac{m}{k}\right)+
N_1^2\left(\frac{m}{k}\right)\right]^{-1}
\ee
which coincides, as one might expect, with the
normalization factor for the massive modes in RS model. In the opposite case
$m\zeta_c\ll 1$ (notice that this automatically implies
$m/k \ll 1$), the expansion
of the Bessel functions in Eqs.\ (\ref{AA}) and (\ref{BB}) yields
\be
C_m^2=\frac{\pi}{(k\zeta_c)^3}\left(1+
\frac{4}{(m\zeta_c)^2(k\zeta_c)^4}\right)^{-1}
\label{21}
\ee

It is now straightforward to calculate the static
gravitational potential between two unit masses
placed on the positive-tension brane at a
distance $r$ from each other.  This potential is
generated by the exchange of the massive modes (cf.\ Refs. 
\cite{Randall:1999vf,Garriga:1999yh} -- recall that the zero mode
here is not localized) 
\be
V(r)=G_5\int_0^\infty~dm~\frac{e^{-mr}}{r} ~\psi_m^2 (z=0)
\label{22}
\ee
It is convenient to divide this integral into two parts,
\be
V(r)=G_5\int_0^{\zeta_c^{-1}}~dm ~\frac{e^{-mr}}{r} ~\psi_m^2(0) +
G_5\int_{\zeta_c^{-1}}^\infty ~dm~\frac{e^{-mr}}{r} ~\psi_m^2(0)
\label{23}
\ee
At $r \gg k^{-1}$, the second term in Eq.\ (\ref{23}) 
is small and it is similar to the contribution of
the continuum modes to the gravitational
potential in  RS model. It gives short distance corrections to Newton's law, 
\be
\Delta V_{short}(r) \sim
\frac{G_5}{kr^3} = \frac{G_N}{r}\cdot \frac{1}{k^2r^2}
\label{24}
\ee
where $G_N=G_5k$ is the four-dimensional Newton constant.

Of  greater interest  is the first term in
Eq.\ (\ref{23}) which dominates at $r \gg k^{-1}$.
Substituting the normalization factor (\ref{21}) into this term, we find
\be
V(r)=\frac{G_5}{r}\int_0^{\zeta_c^{-1}}~dm\frac{\pi}{(k\zeta_c)^3}
\left(1+\frac{4}{(m\zeta_c)^2(k\zeta_c)^4}\right)^{-1}
\frac{4k^2}{\pi^2m^2}e^{-mr}
\ee
This integral is always saturated  at 
$m \aprle r_c^{-1} \ll \zeta_c^{-1}$, where
\be
r_c = \zeta_c (k\zeta_c)^2 \equiv k^{-1} e^{3kz_c}
\label{rrc}
\ee
Therefore, we can extend the integration to infinity and obtain
\bea
V(r) &=& \frac{G_N}{r}\cdot\frac{2}{\pi}\int_0^\infty dx
\frac{e^{-\frac{2r}{r_c}x}}{x^2+1} \label{25} \\
&=& {2G_N\over\pi r} \left [
\mbox{ci} (2r/r_c) \sin (2r/r_c) - \mbox{si} (2r/r_c) \cos (2r/r_c)\right ]
\nonumber
\eea
where $x=mr_c/2$, and $\mbox{ci/si}(t) = -\int_t^\infty {\cos/\sin (u)
\over u}du$ are the sine and cosine integrals. We see that
$V(r)$ behaves in a  peculiar way. At $r\ll r_c$, 
the exponential factor in Eq.\ (\ref{25}) can be set
equal to one and the four-dimensional Newton law is restored,
$V(r)=G_N/r$.
Hence, at intermediate distances,
$ k^{-1} \ll r \ll r_c$, the collection of continuous modes with 
$m \sim r_c^{-1}$ has the same effect as the graviton
bound state in RS model. However,
in the opposite case, $r\gg r_c$, we find
\be
V(r)=\frac{G_Nr_c}{\pi r^2}
\label{5dg}
\ee
which has the form of  ``Newton's law'' of
five-dimensional gravity with a renormalized gravitational constant.

It is clear from Eq.\ (\ref{25}) that at intermediate distances,
$k^{-1} \ll r \ll r_c$, the four-dimensional Newtonian potential
obtains not only short distance corrections, Eq.\ (\ref{24}), 
but also long distance ones, $V(r) = G_N/r + \Delta V_{short}(r)
+ \Delta V_{long}(r)$. The long distance corrections are suppressed by
$r/r_c$, the leading term being
\be
\Delta V_{long}(r)=\frac{G_N}{r}\cdot\frac{r}{r_c}\cdot\frac{4}{\pi}
\left(\ln \frac{2r}{r_c}+{\bf C}-1\right)
\label{26}
\ee
where ${\bf C}$ is the Euler constant. The two types of corrections,
Eqs.\ (\ref{24}) and (\ref{26}), are comparable  at roughly $r\sim
\zeta_c$.
At larger $r$, deviations from the four-dimensional Newton law are
predominantly due to the long-distance effects.

In our scenario, the approximate four-dimensional gravity law is valid over a
finite range of distances. Without strong fine-tuning however,
this range is large, as
required by phenomenology. Indeed, the exponential factor in 
Eq.\ (\ref{rrc}) leads to a very large  $r_c$ even for microscopic
separations, $z_c$, between the branes. As an example, for $k\sim M_{Pl}$
we only require $z_c \sim 50 l_{Pl}$ to have $r_c \sim 10^{28}$ cm,
the present horizon size of the Universe, i.e., with mild assumptions
about $z_c$, the four-dimensional description of gravity is valid from
the Planck to cosmological scales (in this example, long distance
corrections to Newton's gravity law dominate over short distance ones
at $r \aprge \zeta_c \sim 10^{-13}$ cm). 

%ADDITIONAL GRAVITY WAVE STUFF ADDED HERE

An interesting consequence of the incomplete localization of graviton is
the dissipation into the fifth dimension of gravitational waves 
propagating to large distances. Let us consider gravitational waves
generated by a periodic pointlike source on the brane,
$T(x,z)=T({\bf x}) e^{-i\omega t}  \delta (z)$,
where the four-dimensional indices are again omitted.
The gravitational field on the brane
is given by the convolution of the source,
$T({\bf x})$, with the Green's function 
\be
G({\bf x} - {\bf x'}; \omega) =
8\pi G_5\int~d(t'-t)~ G(x,x';z=z'=0) e^{-i\omega (t-t')}
\label{30}
\ee
Here the five-dimensional gravitational constant is included for
convenience of comparison with four-dimensional formulae, and
$G(x,x';z,z')$ is the retarded 
Green's function of the linearized Einstein equations.
The latter
is constructed from the full set of eigenmodes in the
usual way:
\be
G(x,x'; z,z')=\int_{0}^{\infty}dm~\psi_m(z) \psi_m(z') \frac{1}{(2\pi)^4}
\int d^4p\frac{e^{-ipx}}{m^2-p^2-i\epsilon p^0}
\label{31}
\ee
After substitution of (\ref{31}) into (\ref{30}) and
simplifications we get
\be
G({\bf x} - {\bf x'}; \omega) = \frac{2G_5}{r}
\int_0^{\infty}dm~\psi_m^2(0)e^{ip_\omega r}
\label{32}
\ee
where $r=|{\bf x} - {\bf x'}|$,
$p_\omega=\sqrt{\omega^2-m^2}$  when $m<\omega$ and
$p_\omega=i\sqrt{m^2-\omega^2}$  when $m>\omega$.
We see that the gravitational field on the brane
has the form of a superposition of
massive four-dimensional modes. Only modes
with $m<\omega$ are actually radiated, the other ones exponentially fall 
off from the source. Thus, as long as we are interested in gravitational
waves, we can integrate in (\ref{32}) only up to $m=\omega$.

Let us study the case
$r_c^{-1}\ll\omega\ll k$. Then, the main
contribution to (\ref{32}) is given by modes with
$m\sim r_c^{-1}$, whereas modes with larger masses give rise
to
corrections suppressed by $\omega/k$
(cf.(\ref{23}), (\ref{24})). In this region of $m$, the eigenfunctions
$\psi_m$ are given by the
explicit expressions (\ref{18}), (\ref{21}) and $p_\omega$ is approximated
by $\omega-\frac{m^2}{2\omega}$. In this way we get (cf.(\ref{25}))
\be
G({\bf x} - {\bf x'}; \omega)
=\frac{2G_N}{r}e^{i\omega r}\cdot\frac{2}{\pi}
\int_0^{\infty}dx\frac{e^{-i\frac{2r}{\omega r_c^2}x^2}}{1+x^2}
\label{33}
\ee
where again $x=mr_c/2$ and we extended the integration  to
infinity. This integral is expressed in terms of the error function:
\be
G({\bf x} - {\bf x'}; \omega)
=\frac{2G_N}{r} e^{i\omega r} \left ( 1 - 
\erf(\beta) \right ) e^{\beta^2}
\label{34}
\ee
where $\beta=e^{i\frac{\pi}{4}}\sqrt{\frac{2r}{\omega r_c^2}}$.
When $\beta\ll 1$ (that is $r\ll\omega r_c^2$) one has $\erf
(\beta)\approx 0$, and we obtain the usual $1/r$-dependence 
of the gravity wave amplitude on the distance to the source. In 
the opposite case $\beta\gg 1$ ($r\gg \omega r_c^2$), we make use of 
$\erf(\beta) =1-\frac{1}{\sqrt{\pi}\beta} e^{-\beta^2}$ to obtain
\be
G({\bf x} - {\bf x'}; \omega)
=\frac{2G_N}{r^{3/2}}\sqrt{\frac{\omega r_c^2}{2\pi}}
e^{i\omega r-i\frac{\pi}{4}}
\label{35}
\ee
that is, the amplitude is proportional to $r^{-3/2}$. Thus, the 
gravitational waves dissipate into the fifth dimension and from the 
point of view of a four-dimensional observer on the brane, energy of 
gravity waves is not conserved.

This effect becomes considerable after the wave travels the distance of 
order $r\sim \omega r_c^2$ from the source.  Note that at $\omega \gg 
r_c^{-1}$ this distance is much larger than the distance $r_c$ at which 
the violation of four-dimensional Newton's law is appreciable. This 
difference between the two distance scales can in fact be seen to be a 
relativistic effect. The collection of five-dimensional graviton states 
with $m \sim r_c^{-1}$ may be viewed as an RS bound state which becomes 
metastable in our model. The width of this metastable state is of 
order $\Gamma = \Delta m \sim m \sim r_c^{-1}$. In its
own reference frame, the graviton disappears into the fifth dimension
with time scale $\tau \sim \Gamma^{-1} \sim r_c$. This time scale
determines the distance at which four-dimensional gravity of {\it static} 
sources gets modified. On the other hand, when the graviton moves in four 
dimensional space-time with momentum $p \sim \omega$, there is an
additional gamma-factor of order $\gamma \sim \omega/m \sim \omega r_c$.
The graviton therefore remains effectively four-dimensional during a time
of order $\gamma\tau \sim \omega r_c^2$ due to
the relativistic time delay. 

Because of this property, the leaking of the gravity waves into
the extra dimension is negligible at relatively short
wavelengths. However, the corresponding time scale becomes of order
$r_c$ for wavelengths of the same order. This is another manifestation
of our observation that extra dimensions open up at the length scale
$r_c$. 

%AND ENDS HERE

Finally, we point out that in more complicated higher-dimensional models,
the long distance properties of gravitational interactions may be even
more intriguing. Indeed, let us modify the model discussed throughout
this paper by compactifying the fifth dimension to a very large
radius, $z_{*} \gg z_c$. Then the mass spectrum of Kaluza--Klein
gravitons becomes discrete, and the graviton  zero  mode reappears.
However, the spacing between the masses may be tiny, depending on
$z_{*}$. With appropriately chosen $z_{*}$, the five-dimensional
gravity law, Eq.\ (\ref{5dg}), will itself be valid in a finite
interval of distances, and the four-dimensional gravity will again be
restored at largest scales, well above $r_c$. 

We conclude that higher-dimensional theories provide, somewhat
unexpectedly, valid alternatives to four-dimensional 
gravity at {\it large
distances}. With hindsight, it is perhaps not surprising that this
happens, since at the very large scale, the anti-de Sitter sandwich
becomes very slim, and spacetime is nearly flat, however, from the
perspective of our putative universe -- the four-dimensional central wall --
this conclusion is not so transparent. It would be worthwhile to explore
such models in the context of cosmology and astrophysics. 
The long-distance phenomena
in our Universe may become a window to microscopic extra dimensions!

We would like to thank Victor Berezin,
Sergei Dubovsky, Dmitry Gorbunov,
Maxim Libanov and Sergei Troitsky
for useful discussions. We would also like to thank Ian Kogan for drawing
our attention to Ref.\ \cite{Kogan:1999}.
R.G.\ and V.R.\ acknowledge the
hospitality of the Isaac Newton Institute for Mathematical Sciences,
where this work has begun.
R.G.\ was supported in part by the Royal Society, and V.R. and S.S by
the Russian Foundation for Basic Research, grant 990218410.

\end{document}